# The poor altmetric performance of publications authored by researchers in mainland China

Xianwen Wang*, Zhichao Fang, Qingchun Li and Xinhui Guo

WISE Lab, Faculty of Humanities and Social Sciences, Dalian University of Technology, Dalian 116085, China.

* Corresponding author.
Website: http://xianwenwang.com/
Email address: xianwenwang@dlut.edu.cn; xwang.dlut@gmail.com

**Abstract:** China's scientific output has risen precipitously over the past decade; it is now the world's second largest producer of scientific papers, behind only the United States. The quality of China's research is also on the rise (Van Noorden, 2016). The online visibility and impact of China's research are also important issues worth exploring. In this study, we investigate the altmetric performance of publications in the field of Biotechnology and Applied Microbiology and published by authors from Chinese affiliations. We find that papers published by those authors from Chinese affiliations have much lower visibility on the social web than articles from other countries, when there is no significant difference for the citations. Fewer of China's publications get tweeted, and those tweeted publications attract less social attention. A geographical analysis of tweeters shows that scholarly articles get most of their social attention from the authors' home countries, a finding that is also confirmed by correlation and regression analysis. This situation, which is unfavorable for researchers from Chinese affiliations, is caused, in part, by the inaccessibility of mainstream social networking platforms in mainland China.

**Keywords:** Altmetrics, social media, open access, China, Twitter

## Introduction

The prominence of social media has contributed to increased accessibility of scientific information to the public and to increased academic communication among researchers (Hurd, 2013; Virginia et al., 2011). There are several reasons for the frequent use of social media in scientific exchange, such as higher spread efficiency, a wider range of readers, lower cost, and stronger interactivity (Alotaibi et al., 2015; Bornmann, 2015; Wu, 2014). Social media are supposed to break the organizational boundaries of academic communications and promote the dissemination of scientific information in a more open and equal environment. Through the mainstream international social web, on platforms such as Twitter and Facebook, scholarly articles are widely shared and discussed by people all over the world—except for a few countries and regions. Although it is the world's second-largest producer of scientific papers, mainland China is one of these countries that lie outside the mainstream international social web.

In this study, we focus on the relation between the authors' and social media users' geographical locations at the country level. Mainland China was chosen as a specific research object because in mainland China, international mainstream social media platforms, including Facebook and Twitter, are unavailable. Therefore, neither Chinese researchers nor the Chinese public can participate in the mainstream social web that altmetrics track. China's two biggest native social media platforms, Sina Weibo and WeChat, meet the public's requirements for information, and Chinese Internet users do share and discuss scholarly articles on Sina Weibo and WeChat Moments. In 2014, altmetric.com announced that the number of commented and forwarded articles in Sina Weibo will be included in altmetric scores. Nevertheless, the absence of Chinese voices in international social media cannot be remedied by participation in China's native platforms, because these platforms are limited to particular regions.

This study will first present an overall analysis of how social media users' shared native countries affect scholarly articles' altmetric performance. Then, the absence of Chinese on international mainstream social media platforms, and the influence this absence has on scholarly communication, will be discussed.

**Literature review**

As social media becomes an increasingly popular means of sharing scientific publications, altmetrics, which are defined as the creation and study of new metrics based on the social web and tools for analyzing and informing scholarship, have emerged as hot topics in the era of scientometrics 2.0 (Priem et al., 2010; Priem & Hemminger, 2010). Altmetrics measure the impact of research results from various aspects and give a full picture of how research products have influenced conversation, thought, and behavior (Haustein et al., 2013; Piwowar, 2013). The Altmetric Attention Score, which represents a weighted approximation of all the attention altmetric.com has picked up for a research output (Gumpenberger et al., 2016), has been widely embedded in the article-level metrics pages of many publishers and journals.

Given the openness of the Internet, the opportunities for research output to spread on the social web seem to be fair, and the altmetrics evaluations seem to be objective. However, just as geopolitical location, cultural relations, and language shape authors' citation preferences (Schubert & Glänzel, 2006), many factors affect the preferences of social media users. For example, in previous studies of how scholarly articles spread on the social web, academic accounts were found to be more active than ordinary private accounts (Forkosh-Baruch & Hershkovitz, 2012), topical papers were more likely to spread quickly, and news media had a strong impact on the popularity of a scholarly article on Twitter and Facebook (Papworth et al., 2015).

Therefore, the factors that significantly influence the social buzz a scientific publication receives should be attended to. However, few studies have investigated the impact of author nationality on altmetric performance. Moreover, we should be aware that some countries or regions are ignored when research concentrates on mainstream social network on the global level; one of these countries is mainland China.

In the last decade, China has rapidly improved both the quantity and quality of its academic publishing (Van Noorden, 2016). China is now the world's second largest producer of scientific papers, after the United States (Wang, 2016). Scholarly articles published by Chinese authors can be found almost in any journal. Is China's research equally visible on the social web?

In this study, we ask the following research questions: First, how do scholarly publications produced by authors from affiliations in mainland China perform in altmetrics? Second, how much does interaction with social web users from the author's home country affect each article's overall altmetric performance? Finally, does the absence of Chinese people on the international mainstream social web influence the altmetric performance of Chinese publications?

**Data and methods**

Our research objects are publications in the field of Biotechnology and Applied Microbiology, as classified by Web of Science. There are two reasons for this choice. One is that Biotechnology and Applied Microbiology is one of the most productive and specific subject areas; the other is that this subject overall performs well in altmetrics.

The publication data are harvested from Web of Science, while the altmetrics data are from altmetric.com. Because of the open-access advantage, open-access articles are dominant in gaining social media attention (Wang et al., 2015). To avoid errors caused by different access types, all sample articles were chosen from open access journals; these journals have higher visibility and accessibility via social media than non-open access publications, which increases the prospect of public consumption and engagement (Mounce, 2013). The publication data for 6,076 articles in the field of Biotechnology and Applied Microbiology, published in open-access journals in 2015, were retrieved from Web of Science. Using DOIs from the downloaded Web of Science records, we collected the altmetric data for the 6,076 records from altmetric.com, using their API. The altmetric data include the Altmetric Attention Score and number of tweeters, which are used to measure social buzz about the articles in the dataset. The Altmetric Attention Score is a weighted count of the amount of attention altmetric.com picked up for a research output; detailed data sources and weightings of the Altmetric Attention Score can be found at https://help.altmetric.com/support/solutions/articles/6000060969-how-is-the-altmetric-score-calculated. All the data are processed and parsed into a SQL server database for analysis. The final dataset includes the 6,076 identified papers, their Altmetric Attention Scores, the tweeted shares of the papers, and each paper's citations (if any).

In this study, we determine authors' locations based on their institutional affiliation. For example, if the author's institution is located in mainland China, that author is defined as an "author in mainland China". Here we use "author in mainland China" instead of "author from mainland China", because only the authors from institutions located in China are considered, and those Chinese authors with affiliations from other countries are excluded. Accordingly, there are two ways in which the 6,076 articles of the dataset are classified into two groups. In the first method, articles are divided based on the location of *all* authors' affiliations: either all article authors are in mainland China or no article authors are in mainland China. In the second classification method, articles are grouped instead by the location of the first or corresponding author's affiliation.

Table 1. Two data grouping methods based on the location of authors' affiliations

| Grouping method | Group | Publications | Publications with Altmetric Attention Score (percentage) | Publications that have been tweeted at least once |
| --- | --- | --- | --- | --- |
| I | All authors in mainland China | 1195 | 636 (53.22%) | 553 (46.28%) |
| | All authors NOT in mainland China | 4420 | 2671 (60.34%) | 2563 (57.99%) |
| II | First or corresponding author in mainland China | 1364 | 735 (53.89%) | 648 (47.51%) |
| | First or corresponding author NOT in mainland China | 4483 | 2718 (60.63%) | 2610 (58.22%) |

As Table 1 shows, for the group in which all authors are in mainland China, 53.22% of articles have Altmetric Attention Scores and 46.28% of articles have been tweeted at least one time; both of these metrics are lower than those seen in the group without authors in mainland China, 60.34% (Altmetric Attention Score) and 57.99% (tweeters). When articles are divided instead by location of the first or corresponding authors' affiliations, the results are similar: much fewer Chinese publications have Altmetric Attention Scores and Chinese publications are tweeted much less often than articles from other countries.

**Results**

*Top papers as measured by Altmetric*

3,530 of the total 6,076 papers have an Altmetric Attention Score greater than 0. Only one paper has an Altmetric Attention Score greater than 100; this paper reached 133. Here we analyze the top 100 most popular papers as measured by altmetric.com. Thirty-one of the top 100 have first authors in the United States, eleven papers have first authors in the United Kingdom, and seven papers have first authors in mainland China. As Table 2 shows, if we consider the total publications with first authors from each country and calculate the proportion, only 0.48% of first-authored papers produced by Chinese affiliations are ranked in the top 100—much lower than most other countries (e.g., the proportion of the United States is 3.72% and that of the United Kingdom 7.01%).

Table 2. Percentage of top papers by country (based on the location of first author's affiliation)

| Country/region | Top 100 papers | All first-authored papers | Proportion |
| --- | --- | --- | --- |
| USA | 31 | 834 | 3.72% |
| UK | 11 | 157 | 7.01% |
| mainland China | 7 | 1455 | 0.48% |
| Germany | 6 | 298 | 2.01% |
| France | 5 | 152 | 3.29% |
| Canada | 4 | 105 | 3.81% |
| South Korea | 4 | 207 | 1.93% |
| Australia | 3 | 111 | 2.70% |
| Denmark | 3 | 47 | 6.38% |
| Italy | 3 | 263 | 1.14% |
| Japan | 3 | 220 | 1.36% |
| Netherlands | 3 | 104 | 2.88% |
| Czech Republic | 2 | 42 | 4.76% |
| India | 2 | 166 | 1.20% |
| Spain | 2 | 162 | 1.23% |
| Austria | 1 | 52 | 1.92% |
| Belgium | 1 | 76 | 1.32% |
| Brazil | 1 | 244 | 0.41% |
| Chile | 1 | 23 | 4.35% |
| Finland | 1 | 35 | 2.86% |
| Ireland | 1 | 16 | 6.25% |
| Poland | 1 | 145 | 0.69% |
| Saudi Arabia | 1 | 55 | 1.82% |
| Slovakia | 1 | 6 | 16.67% |
| Sweden | 1 | 73 | 1.37% |
| Taiwan | 1 | 188 | 0.53% |

For corresponding authors, the results turn out the same: for 1,455 papers with corresponding authors from Chinese affiliations, the proportion of top 100 papers is still only 0.48%, while the United States has 3.78%, the United Kingdom has 6.79%, France has 3.27%, Canada has 3.60%, and Germany has 1.32% (as shown in Table 3).

Table 3. Percentage of top papers by country (based on the location of corresponding author's affiliation)

| Country/region | Top 100 papers | All corresponding authored papers | Proportion |
|---|---|---|---|
| USA | 33 | 873 | 3.78% |
| UK | 11 | 162 | 6.79% |
| Peoples R China | 7 | 1455 | 0.48% |
| France | 5 | 153 | 3.27% |
| Canada | 4 | 111 | 3.60% |
| Germany | 4 | 304 | 1.32% |
| Australia | 3 | 117 | 2.56% |
| Denmark | 3 | 50 | 6.00% |
| Italy | 3 | 258 | 1.16% |
| Japan | 3 | 223 | 1.35% |
| Netherlands | 3 | 97 | 3.09% |
| South Korea | 3 | 207 | 1.45% |
| Czech Republic | 2 | 41 | 4.88% |
| India | 2 | 160 | 1.25% |
| Spain | 2 | 158 | 1.27% |
| Austria | 1 | 50 | 2.00% |
| Belgium | 1 | 77 | 1.30% |
| Brazil | 1 | 245 | 0.41% |
| Chile | 1 | 24 | 4.17% |
| Finland | 1 | 41 | 2.44% |
| Ireland | 1 | 15 | 6.67% |
| Poland | 1 | 145 | 0.69% |
| Saudi Arabia | 1 | 60 | 1.67% |
| Slovakia | 1 | 5 | 20.00% |
| Sweden | 1 | 73 | 1.37% |
| Switzerland | 1 | 56 | 1.79% |
| Taiwan | 1 | 187 | 0.53% |

*Comparison of average value*

Figure 1 compares the average Altmetric Attention Scores and tweeters between the two groups divided using grouping method I (all authors in mainland China/no authors in mainland China). Here we calculate the mean value instead of median value, because many articles have never been tweeted, and many articles have been tweeted by only one or two tweeters; the median value could not, therefore, reflect the difference between the two groups as accurately as the mean value does. The mean Altmetric Attention Score for the group with all authors in mainland China is 1.23, much lower than in the other group with no authors in mainland China, which is 2.41. The same result is reflected in the average number of tweeters. The articles whose authors are all in mainland China have an average of 1.08 tweeters, while the other group (no authors in mainland China) has an average of 2.58 tweeters. The difference between the groups is smaller for citation metrics.

Grouping method II, which groups articles according to the country of the first or corresponding author's affiliation, also returns differences. For articles whose first or corresponding author is in mainland China, the mean Altmetric Attention Score is 1.26, the mean number of tweeters is 1.16, and the mean number of citations is 0.56; for the other group, whose first or corresponding authors are not in mainland China, the mean Altmetric Attention Score is 2.45, the mean number of tweeters is 2.63, and the mean number of citations is 0.64. For the Altmetric Attention Score and tweeters, the difference between the two groups is huge; for citation, the difference is very small.

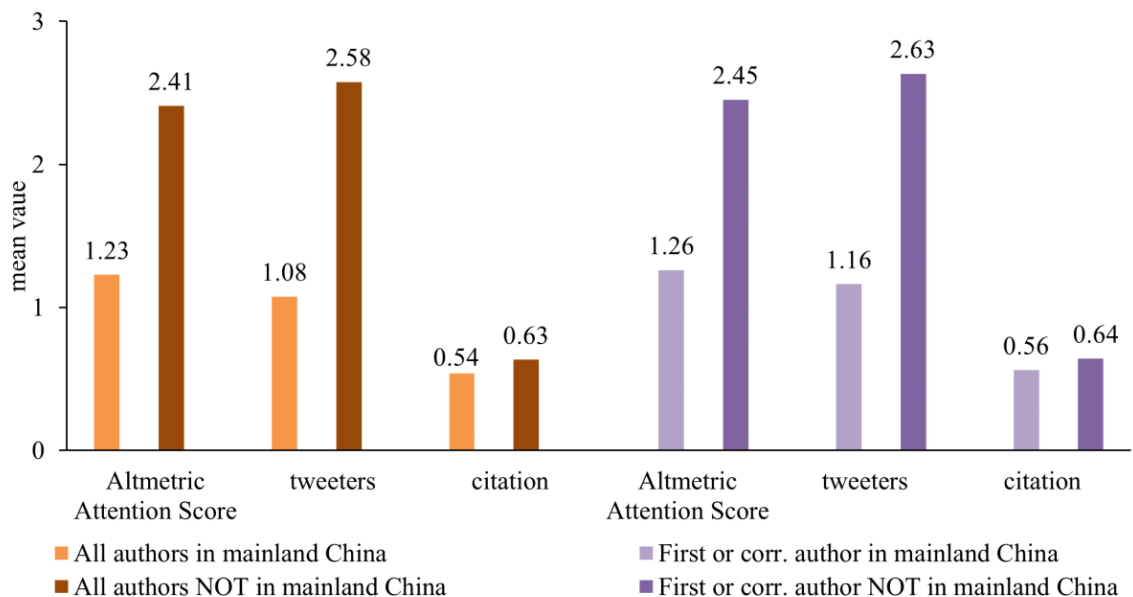

Figure 1. Comparison of mean Altmetric Attention Scores and tweeters

Although the number of citations for articles from mainland China is slightly lower than that of the articles from other countries, the difference in citation numbers is much smaller than the difference in altmetrics (which include both Altmetric Attention Score and the number of times the article was tweeted). Although articles from mainland China have a small disadvantage in citation impact, this is not enough to explain their poor performance on social media.

*Nonparametric test*

In order to statistically compare the altmetric performance of articles from China to that of articles from other countries, we used the 2-sample Kolmogorov-Smirnov nonparametric test, performed on IBM SPSS Statistics 23, to evaluate whether the two samples come from the same distribution. The two samples include the group with articles from China and the group with articles from other countries, as defined in Table 1. The null hypothesis is that both groups are sampled from populations with identical distributions; we tested for any violation of that null hypothesis, including different medians, different variances, or different distributions.

As Table 4 (classified by whether *all* authors are in mainland China) and Table 5 (classified by whether first or corresponding author is in mainland China) show, neither the Altmetric Attention Score nor the number of tweeters have the same distributional function across the two samples (Kolmogorov-Smirnov Z = 4.327/4.297/4.922/4.826, p = .000); we can conclude that the two groups were sampled from populations with different distributions. However, the comparison of citation rates for both groups have p-values far greater than 0.05, indicating that there is no significant difference in citations between the group of articles from China and the group of articles from other countries. Although Chinese publications have a slightly lower mean number of citations (as shown in Figure 1), the difference is not significant. The nonparametric test excludes the possibility that the lower social media exposure of Chinese publications is a product of their lower citation impact.

Table 4. 2-Sample Kolmogorov-Smirnov Test of groups classified by whether *all* authors are in mainland China

|  | Kolmogorov-Smirnov Z | P-value |
|---|---|---|
| Altmetric Attention Score | 4.327 | .000 |
| Tweeter | 4.297 | .000 |
| Citation | .685 | .736 |

Table 5 2-Sample Kolmogorov-Smirnov Test of two groups classified by whether first or corresponding author is in mainland China

|  | Kolmogorov-Smirnov Z | P-value |
|---|---|---|
| Altmetric Attention Score | 4.922 | .000 |
| Tweeter | 4.826 | .000 |
| Citation | .660 | .776 |

Both the average values and the nonparametric tests indicate that publications whose authors from Chinese affiliations have the relatively poor altmetric performance (based on Altmetric Attention Score and tweeters).

*Geographical analysis of tweeters*

On the altmetric.com details page, the Twitter Demographics shows the geolocation data collected from the profiles of tweeters who shared the paper (altmetric.com geolocates users based on their profile information).

In our dataset, 3,334 articles were tweeted by 13,712 tweeters (only eight papers were shared ten times on Sina Weibo); the locations of 6,853 tweeters are unknown. In Table 6, we list the countries that have had articles tweeted by more than 100 tweeters. The United States, the United Kingdom, and France have far more tweeters than any other countries.

Table 6. Countries ranked by number of articles tweeted by more than 100 tweeters

| Rank | Country | Tweeters |
|---|---|---|
| 1 | USA | 1855 |
| 2 | UK | 1503 |
| 3 | France | 990 |
| 4 | Spain | 260 |
| 5 | Australia | 208 |
| 6 | Canada | 208 |
| 7 | Italy | 197 |
| 8 | India | 166 |
| 9 | Norway | 139 |
| 10 | Germany | 139 |
| 11 | Japan | 133 |
| 16 | China | 60 |

We examine articles whose first or corresponding author is in one of the twelve countries listed in Table 6 (which includes the top eleven countries with articles tweeted by more than 100 tweeters, based on geolocations of the tweeters who tweeted the publications, as well as China). Only articles whose first or corresponding author has a unique affiliation from mainland China' affiliation are considered to be Chinese articles; if the first or corresponding author has multiple affiliations and one or more is not in mainland China, that article was excluded from the dataset.

In Figure 2, we calculate the percentage of tweeters from different countries. Each stacked bar represents the percentage of tweeters from one country. The bar length is decided by the percentage: the closer to the y axis, the more tweeters are from that country. Most of the data markers are in grey color; the bar highlighted in red represents the tweeters from the home country where the author's affiliation located in.

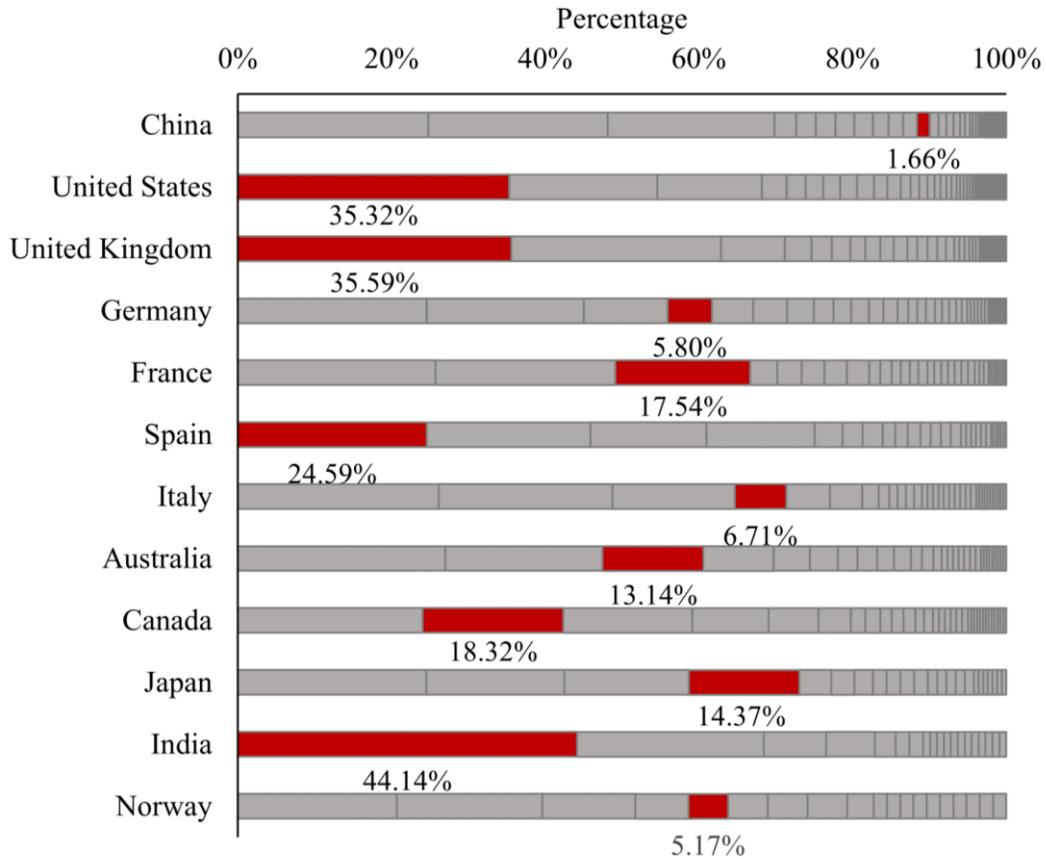

Figure 2. Tweeters from other countries

As Figure 2 shows, articles whose first or corresponding author is in the United States, the United Kingdom, Spain, and India have tweeters from the author's home country contributing the most tweeters about those articles. Articles published by authors from France, Canada, and Australia have an advantage as well; tweeters from these authors' home countries rank second or third among all countries. For articles published by German, Italian, and Japanese authors, the authors' home country ranks fifth in Tweeters among all countries. However, tweeters from China are rather rare. For most articles, tweeters from the United States and the United Kingdom contributed the most tweeters. Tweeters from the United States account for 35.32% of tweeters about articles published by US researchers. For the United Kingdom, the percentage is 35.59%, about the same. Tweeters from India account for as much as 44.14% of the tweeters about articles by Indian authors. For Spain, France, Australia, Canada, and Japan, the percentages are between 13% and 25%. For Germany, Italy, and Norway, the percentages are between 5% and 7%. However, for China, the percentage is only 1.66%, which is much lower than other countries, especially India, the United States, the United Kingdom, and Spain.

Research from Germany, France, Italy, Australia, Japan, and Norway is most often tweeted, not by tweeters from the authors' home countries, but by users in the United States and the United Kingdom. This may be because Twitter has the highest level of penetration in those two countries, meaning they have larger numbers of tweeters who actively participate in sharing and discussing scientific output on the social web.

*Regression analysis*

To better understand what role people in author's home country play in diffusing scholarly articles on the social web, we conducted correlation analysis and regression analysis. Here we choose the 1,683 articles that were tweeted at least one time and that have first or corresponding authors from the affiliations in any countries listed in Figure 2 except China (the United States, the United Kingdom, Germany, France, Spain, Italy, Australia, Canada, Japan, India, and Norway). For each article, we calculate the total tweeters who tweeted the article (dependent variable) and the number of tweeters from the first or corresponding author's home country who tweeted the article (independent variable). The regression analysis is conducted based on individual article. Table 7 summarizes the descriptive statistics and correlation analysis results.

Unlike other countries, few tweeters who tweet Chinese publications are from China; because people in mainland China have no access to Twitter, the Twitter data for Chinese publications differs from that other countries' publications, which could bias the results. We therefore excluded data from China in the regression analysis.

In this part, we would like to examine the effect of tweeters from the author's home country on the diffusion of articles on social media. Since the data are positively skewed, we conducted Spearman correlation analysis. As can be seen in Table 7, the number of tweeters from the country of first or corresponding author's affiliation who tweet an article is positively and significantly correlated with the number of total tweeters who tweet an article; the correlation coefficient r is 0.548 ($p<0.01$).

Table 7. Summary statistics, correlations and results from the regression analysis

| Variable | Mean | Standard deviation | Correlation with total tweeters | Regression weights (Linear regression with logarithmic transformations) | | |
|---|---|---|---|---|---|---|
| | | | | b | β | $R^2$ |
| Total tweeters | 5.22 | 7.655 | | | | |
| Tweeters from the country of first or corresponding author's affiliation | .71 | 1.944 | .548** | .799*** | .650 | .422 |

Note: * $p < .05$ ** $p < .01$ ***$p < .001$

Because the data is positively skewed, linear regression of original variables is not appropriate. However, variables with logarithmic transformations obey normal distribution. Therefore, in this study, variables are log-transformed using natural logarithms.

Correlation and linear regression analysis with logarithmic transformations are conducted to examine the relationship between the total tweeters and tweeters from the first or corresponding author's country. Table 7 summarizes the descriptive statistics and analysis results. The total number of tweeters who tweet an article are positively and significantly correlated with the number of tweeters from the first or corresponding author's country who tweet an article, r=.548, p < .01. This indicates that articles tweeted by more tweeters from the author's home country tend to be tweeted by more total tweeters.

The results of the regression indicate that the independent variable explains 42.2% of the variance ($R^2$ =.422, $F(1, 1681) = 370.235$, $p < .001$). It is found that the number of tweeters from the first or corresponding author's country who tweet an article significantly predicts the total number of tweeters who tweet that article ($\beta = .799$, $p < .001$). As shown in Table 7, tweeters from the first or corresponding author's country have significant positive regression weights, indicating that articles tweeted by more tweeters from the first or corresponding author's country are expected to be tweeted by more tweeters in total. Social buzz from the author's country contributes a lot to broad diffusion of scholarly articles on social media.

**Conclusion**

Even though the scholarly impact of China's output has been improved significantly, and its authors feature on around one-fifth of the world's most-cited papers now (Van Noorden, 2016), by contrast, the social impact of China's publication is unimpressive. As to our first research question, without significant difference for the citations, articles published by authors in mainland China have relative poor performance on altmetrics in two dimensions. First, compared to other countries, fewer publications by authors in mainland China are tweeted—only about 46%. In contrast, 58% of articles by non-authors in mainland China are tweeted at least once. Secondly, China's publications attract less social media attention. Articles by authors in mainland China have an average Altmetric Attention Score of about 2.3, and are tweeted by not more than 3 tweeters. Articles by authors in other countries have an average Altmetric Attention Score of about 4 and are tweeted by an average of 4.5 tweeters. The results of nonparametric tests also indicate that China's publications perform relatively poorly on altmetrics (including both the Altmetric Attention Score and article tweeters).

For the second research question, the geographical analysis of tweeters shows that Twitter accounts from an author's home country contribute significantly in diffusing scholarly articles on the social web, as indicated by the correlation and regression analyses. Articles published by authors in the United States, the United Kingdom, Spain, and India have large blocks of tweeters from the author's home country. In addition, the United States and the United Kingdom have the highest levels of Twitter penetration; this could explain why tweeters from the United States and the United Kingdom contributed the most tweeters for most articles in the dataset (see Figure 2). However, publications by authors from Chinese affiliations are tweeted seldom by Chinese tweeters, because of China's absence from the main social media platforms, Twitter and Facebook. Using altmetrics indicators to evaluate publications is thus unfair for authors from Chinese affiliations.

The absence of Chinese people on the international mainstream social web does indeed negatively influence the altmetric performance of Chinese publications. The reasons for China's lack of social media presence include limitations of language and internet, yet the inaccessibility of mainstream social media platforms is the greatest barrier for the Chinese general public and researchers, and this inaccessibility has produced the extended downturn in Chinese publications' altmetric performance.

**Discussion and limitations**

There are several possible reasons why social media users are more inclined to spread and discuss the scientific publications produced by those authors from their home countries' institutions. And the first point is that people may pay more attention to the progress in science made by their home countries and have more willing to spread it.

In addition, peers from the same institutions/regions/countries with the authors may be another important factor that improves the diffusion on social web. Generally speaking, the relationship between researchers from the same country is much stronger than it between researchers from different countries. There are more frequent academic exchanges and collaborations among researchers from the same country. Therefore, researchers may be more likely to spread and discuss the publications produced by those researchers they know or even familiar with in both formal (citation or academic conference) and informal (social network discussion) ways.

Unfortunately, all of these factors are unfair for Chinese publications. Neither the public nor researchers in mainland China have the chances in participating in the discussion on international social media platforms, which leads to the inactivity of Chinese publications to some extent.

Here are a few additional thoughts about the results. Unlike traditional evaluation methods, the altmetrics database is inherently unfair; universal access does not exist for any social media platform. The connection between altmetric performance and the geolocation of author's affiliation is so strong that the accuracy and credibility of altmetrics on the global level should be carefully evaluated—especially when the evaluation objects come from mainland China, the world's second-largest producer of scientific publications.

There are some limitations in this study. First, the dataset used in this study is not large; only 6,076 papers, all published in a single year, are included. Second, the disciplinary coverage is limited, since the dataset deals only with publications in one discipline,

Biotechnology and Applied Microbiology. Third, the geographical analysis data and the regression analysis include only those tweeters with identifiable locations, and exclude tweeters with unknown locations. Fourth, Twitter is one of the most important sources of Altmetric Attention Score, but it could not perfectly represent all aspects of altmetric performance. It should also be noted that there are many other sources to evaluate the altmetric performance of scientific publications, such as Facebook, Blog, LinkedIn, etc. In this study, due to the availability of geographical data, we only take Twitter as an example to investigate users' geographical preferences. If more geographical data were provided by any other sources in the future, we would continue to extend our conclusion on a broader basis. Finally, the Altmetric Attention Score used in this study to reflect altmetrics impact has some potential pitfalls, including possible interdependence of components, weights chosen practically arbitrarily, and information loss from the linearization of the original multi-dimensional space (Gumpenberger et al., 2016).

**Acknowledgements**

The work was supported by the project of "National Natural Science Foundation of China" (61301227, 71673038), the project of "Growth Plan of Distinguished Young Scholar in Liaoning Province" (WJQ2014009) and the project of "the Fundamental Research Funds for the Central Universities" (DUT15YQ111).